\documentclass[twocolumn,prb]{revtex4}
\usepackage[colorlinks,bookmarks=false,citecolor=blue,linkcolor=red,urlcolor=blue]{hyperref}
\usepackage{graphicx,epstopdf,color}
\usepackage{amsfonts}
\usepackage{amsmath,amssymb,mathrsfs}
\usepackage{bm}  
\usepackage{ucs} 
\usepackage[utf8x]{inputenc} 
\usepackage[T1]{fontenc}
\usepackage[final,inline]{showlabels} 
\usepackage{mdframed,enumitem}
\usepackage{dcolumn}
\usepackage{tikz}
\usetikzlibrary{calc,arrows,decorations.pathreplacing,decorations.markings}

\newcommand{\bs}[1]{\boldsymbol{#1}}

\newcommand{\braket}[2]{\left\langle #1 | #2 \right\rangle}

\newcommand{\bra}[1]{\left\langle#1\right|}
\newcommand{\ket}[1]{\left|#1\right\rangle}

\newcommand{\bignorm}[1]{\bigl\Vert #1\bigr\Vert}

\renewcommand{\i}{\text{i}}
\newcommand{\f}{\text{f}}

\newcommand{\x}{\text{x}}

\newcommand{\z}{\text{z}}
\newcommand{\M}{\text{M}}
\renewcommand{\S}{\text{S}}

\newcommand{\up}{\uparrow}
\newcommand{\dw}{\downarrow}

\newlength{\ytlength}
\allowdisplaybreaks[1]
\begin{document}
\title{Sidney Coleman's Dirac Lecture ``Quantum Mechanics in Your Face''}
\author{Sidney Coleman\\[6pt]  Transcript and slides edited by Martin Greiter}
\affiliation{Institute for Theoretical Physics, University of
  Würzburg, Am Hubland, 97074 Würzburg, Germany}

\pagestyle{plain}
\date{\today}

\begin{abstract}
This is a write-up of Sidney Coleman's classic lecture first given as a Dirac Lecture at Cambridge University and later recorded when repeated at the New England sectional meeting of the American Physical Society (April 9, 1994).  My sources have been this recording\cite{coleman} and a copy of the slides Sidney send to me after he gave the lecture as a Physics Colloquium at Stanford University some time between 1995 and 1998.  To preserve both the scientific content and most of the charm, I have kept the editing to a minimum, but did add a bibliography containing the references Sidney mentioned.---MG
\end{abstract}

\maketitle

This lecture has a history.
It's essentially a rerun of a lecture I gave as the Dirac Lecture at Cambridge University a little under a year ago.

There's a story there.
I had been asked to give this lecture several years ago, when it was two years in the future. And of course when someone asks you to do something two years in the future, that's never. You'll always say yes.

And when the time came I got a communication from Peter Goddard at St.~John's College, who was running the operation. He said: ``What do you want to talk about?'' I said: ``Who's the audience?'' And he said: ``Oh, it's pretty mixed---you'll get physics graduate students, physics undergraduates, people from chemistry and philosophy and mathematics.'' And I thought: ``Hmmm, these are not the people whom to address on the subject of non-abelian quantum hair on black holes'', which was what I was working on at the moment.
\begin{figure}[htb] 
\begin{mdframed}
  \vspace{5pt}\hfill 1\\[-15pt]
  \begin{center}
  \begin{obeylines}{\bf
  Quantum Mechanics in Your Face\\[8pt]
  It's Quantum Mechanics, Stupid\\[8pt]
  And Now for Something Completely
  Different: Quantum Reality\\[8pt]
  }
  {Quantum Mechanics with the Gloves Off}\\[10pt]
  \vspace{8pt}
  \end{obeylines}
\end{center}
\end{mdframed}
\end{figure}

So I said: ``Look, I've always been interested in giving a lecture on quantum mechanics---what a strange thing it is, and exactly what strange thing it is---do you think such a lecture would be suitable?'' And he said: ``Yes, give us a clever title!'' So I emailed back: \emph{Quantum Mechanics in Your Face}, because I wanted to really confront people with quantum mechanics. [Coleman places Slide 1 on the projector, with all the titles except for the first one covered up.  He then uncovers the titles as he presents them.] And Peter said: ``No good''. He said a British audience would not understand the locution and indeed might think it was obscene.

``All to the good!'' I said. But he was adamant.

So, since one of the themes of the proposed lecture was that a lot of confusion arises because people keep trying to think of quantum mechanics as classical mechanics, I suggested this alternative title \emph{It's Quantum Mechanics, Stupid}. And he said---I have all of this on disks: this is a true story---``Nope; a British audience wouldn't get it; too American''. So I said: ``Well, all right, if you want something British: \emph{And Now for Something Completely Different: Quantum Reality}.'' He said: ``Too facetious''.

So finally we settled on the title \emph{Quantum Mechanics with the Gloves Off}, which, as you can see, is a little wimpier than the others.

But now I'm back in the land of free speech, so the title of the talk is \emph{Quantum Mechanics in Your Face}.

The talk will fall into three parts.

\begin{figure}[htb] 
\begin{mdframed}
  \vspace{5pt}\hfill 2\\[-18pt]
  \begin{center}
    \textbf{Outline}
  \end{center}
  \begin{raggedright}
  \begin{enumerate}[label=(\arabic*)]
  \item A quick review of (vernacular) quantum \\ mechanics
  \item Better than Bell: the GHZM effect
  \item The return of Schrödinger's cat
  \end{enumerate}
    There is no representation, expressed or implied, that any part of this lecture is original\footnote{or that any account is taken of classical or quantum gravity.}
  \end{raggedright}
  \vspace{6pt}
\end{mdframed}
\end{figure}

There will be a preliminary where I give a quick review of quantum mechanics---I would say 
the Copenhagen interpretation, or the interpretation in somebody's textbook, but it's not really that---it's looser and more sloppy.

Architects and architectural historians, when they're discussing kinds of buildings that were being built in a certain place in a certain time, but aren't in any particular well defined style, but just what builders threw up in the United States in ca.\ 1948---they call it ``vernacular architecture''. This will be a quick review of vernacular quantum mechanics. That's more to establish notation, and make sure we're all on the same wavelength.

Then the two main parts of the lecture will be, firstly, a review of a pedagogical improvement on John Bell's famous analysis of hidden variables in quantum mechanics\cite{bell64physics195,bell66rmp447}. It is easier to explain than Bell's original argument, and deserves to be widely publicized. It was built by David Mermin\cite{mermin90ajp731,mermin90pt9}. He's the ``M'' out of some earlier work by Greenberger, Horn, and Zeilinger\cite{greenberger-89proc,greenberger-90ajp1131}.

In the second [main] part of the lecture I will turn to the much vexed question sometimes called ``the interpretation of quantum mechanics'', although, as I will argue, that's really a bad name for it. I want to stress that I have made no original contributions to this subject. There is nothing I will say in this lecture, with the exception of the carefully prepared spontaneous jokes---that was one of them---that cannot be found in the literature.

Of course, such is the nature of the subject that there is nothing I will say where the contradiction cannot also be found in the literature. So I claim a measure of responsibility, if no credit---the reverse of the usual scholarly procedure.

I will stick strictly to quantum mechanics in flat space and not worry about either classical or quantum gravity. We will have problems enough keeping these things straight there without worrying about what happens when the geometry of space-time is itself a quantum variable.

\begin{figure}[htb] 
\begin{mdframed}
  \vspace{5pt}\hfill 3\\[-18pt]
  \begin{center}
    \textbf{(1) Some Things Everyone Knows}\\[4pt]
    (Even if not everyone believes them)
  \end{center}
  \begin{enumerate}[label=(\roman*)]
  \item The state of a physical system at a fixed time is a vector in a Hilbert space, $\ket{\psi}$, normalized such that $\braket{\psi}{\psi}=1$.
  \item It evolves in time according to
    \begin{align}
      \nonumber
      \i\frac{\partial}{\partial t}\ket{\psi} = H\ket{\psi}
    \end{align}
    where $H$ is ``the Hamiltonian'', some self-adjoint linear operator.
  \end{enumerate}
  \vspace{6pt}
\end{mdframed}
\end{figure}

Now to begin with, the very quick review. These transparencies are going to go by extremely fast. A pointer might be handy.... I'm always worried about these things: I'll point them the wrong way and zap a member of the audience. 

The state of a physical system at a fixed time is a vector in Hilbert space. Following Dirac we call it $\psi$. We normalize it to unit norm. It evolves in time according to the Schrödinger equation, where the Hamiltonian is some self-adjoint linear operator---a simple one if we're talking about a single atom, and a complicated one if we're talking about a quantum field theory.

Now if there is anyone who has any questions about the material on the screen at this moment, please leave the auditorium, because you won't be able to understand anything else in the lecture.

\begin{figure}[htb] 
\begin{mdframed}
  \vspace{5pt}\hfill 4\\[-18pt]
  \begin{raggedright}
  \begin{enumerate}[label=(\roman*)]
    \setcounter{enumi}{2}
  \item Some (maybe all) self-adjoint operators are ``observables''.  If 
    $\ket{\psi}$ is an eigenstate of the observable $A$ with eigenvalue $a$,
    \begin{align}
      \nonumber
      A\ket{\psi} = a\ket{\psi}
    \end{align}
    then we say ``the value of $A$ is certain to be observed to be
    $a$''.\\[6pt]

    [Strictly speaking, just a definition, but there is an implicit
      promise (c.f. $F=ma$).]
  \end{enumerate}
  \end{raggedright}
  \vspace{6pt}
\end{mdframed}
\end{figure}

Now some, maybe all, self-adjoint operators are ``observables''. If the state is an eigenstate of an observable $A$, with eigenvalue $a$, then we say the value of $A$ is $a$, is certain to be observed to be $a$. Now, strictly speaking, this is just a definition of what I mean by ``observable'' and ``observed'', but of course that's because those words have not occurred on any previous transparency so I can call them what I want. Of course, that's like saying Newton's second law
$F = ma$, as it appears in textbooks on mechanics, is just a definition of what you mean by ``force''. That's true, strictly speaking, but we live in a landscape where there is an implicit promise that when someone writes that down, when they begin talking about particular dynamical systems that they will give laws for the force, and not, say, for some quantity involving the 17th time derivative of the position.

Likewise, the words ``observable'' and ``observed'' have a history before quantum mechanics. People like to say all these things have a meaning in classical mechanics, but really it goes way earlier than classical mechanics. I'm sure the pre-Columbian inhabitants of Massachusetts were capable of saying, in their language, ``I observe a deer'', despite their scanty knowledge of Newtonian mechanics. Indeed, I even suspect that the deer was capable of observing the Native Americans despite its even weaker grasp on action and angle variables.

So there's an implicit promise in here that, when you put the whole theory together and start calculating things, that the words ``observes'' and ``observable'' will correspond to entities that act in the same way as those entities do in the language of everyday speech under the circumstances in which the language of everyday speech is applicable. Now to show that is a long story. It's not something I'm going to focus on here, involving things like the WKB approximation and von Neumann's analysis of an ideal measuring device\cite{Neumann32}, but I just wanted to point out that that's there.

\begin{figure}[htb] 
\begin{mdframed}
  \vspace{5pt}\hfill 5\\[-18pt]
  \begin{raggedright}
  \begin{enumerate}[label=(\roman*)]
    \setcounter{enumi}{3}
  \item Every measurement of $A$ yields one of the eigenvalues of $A$.  The probability of finding a particular eigenvalue, $a$, is 
    \begin{align}
      \nonumber
      \lVert P(A;a)\ket{\psi}\rVert^2
    \end{align}
    where $P(A;a)$ is the projection operator on the subspace of states with eigenvalue $a$. (I assume, for notational simplicity, that $A$ has a discrete spectrum.) If $a$ has been measured, then the state of the system after the measurement is
    \begin{align}
      \nonumber
      \frac{P(A;a)\ket{\psi}}{\lVert P(A;a)\ket{\psi}\rVert}
    \end{align}
    (Much) more about this later.
  \end{enumerate}
  \end{raggedright}
  \vspace{6pt}
\end{mdframed}
\end{figure}

Now we come to the fourth thing:  every measurement that happens when the state $\psi$ is not an eigenstate of the observable yields one of the eigenvalues, with the probability of finding a particular eigenvalue $a$ proportional to the magnitude of the part of the wave function that lies on the subspace of states with eigenvalue $a$. (I'm assuming here just for notational simplicity that the eigenvalue spectrum is discrete.) If $a$ has been measured, then the state of the system after the measurement is just that part of the wave function---all the rest of it has been annihilated. And, of course, it has to be rescaled, or being a quantum field theorist, I suppose I should say renormalized, so it has unit norm again. This is the famous projection postulate.  It is sometimes called ``the reduction of the wave packet''. 

It's very different from the previous three statements I've put on the board because it contradicts one of them: causal time evolution according to Schrödinger's equation.  Schrödinger's equation,
\begin{align}
  \frac{d\ket{\psi}}{dt} = -\i H\ket{\psi},
\end{align}
is totally causal: given the initial wave function---given the initial state of the system---the final state is completely determined. Furthermore, this causality is time reversal invariant: given the final state the initial state is completely determined.

This operation is something other than Schrödinger's equation. It is not deterministic. It is probabilistic. It isn't just that you cannot predict the future from the past. Even when you know the future, you don't know what the past was. If I measure an electron and discover it is an eigenstate of $\sigma_z$ with $\sigma_z = +1$, I have no way of knowing what its initial state was. Maybe it was $\sigma_z = +1$, maybe it was $\sigma_x = +1$, and it turned out that I was in the 50\% probability branch that got the measurement $\sigma_z = +1$.

That ends the preliminary.
\medskip

Before I go into the first of the two main parts of the lecture, the GHZM \cite{greenberger-89proc,mermin90pt9}
analysis, are there any questions about this?

\begin{figure}[htb] 
\begin{mdframed}
  \vspace{5pt}\hfill 6\\[-18pt]
  \begin{center}
    \textbf{(2) Credits for the next part}
  \end{center}
  \begin{raggedright}
  \begin{enumerate}[label={[\roman*]}]
    \item A. Einstein, B. Podolsky, and N. Rosen,\\ 
      Phys. Rev. 47, (1935) 777. 
    \item J. S. Bell, 
      Rev. Mod. Phys. 38, 447 (1966);\\
      ---, 
      Physics 1 (1964) 195.
    \item N.D. Mermin, 
      Physics Today 38, April 1985, p. 38.

    \item D.M. Greenberger, M.A. Horne, A. Shimony, and A. Zeilinger, 
      Am. J. Phys. 58 (1990) 1131.
      
    \item N.D. Mermin, 
      Am. J. Phys. 58 (1990) 731;\\
      ---, 
      Physics Today 43, June 1990, p. 9.
  \end{enumerate}
  \end{raggedright}
  \vspace{6pt}
\end{mdframed}
\end{figure}

I will in the second part return 
to a critical analysis of the ``reduction of the wave packet'', but for the first part of this lecture I'd like to take it as given. Now there's references, but actually I call them credits, because I noticed nobody ever writes down the references. It is just here to avoid the speaker being sued. This whole analysis, as everyone knows, starts with the work of Einstein, Rosen, and Podolsky\cite{einstein-35pr777}, which sat around as an irritant for some years, until John Bell\cite{bell64physics195,bell66rmp447}, picking up an idea from David Bohm\cite{Bohm51}, was able to turn it into a conclusive argument against hidden variables.

A pedagogical improvement was made by David Mermin\cite{mermin85pt38} who, at least to my mind, really clarified what was going on in Bell's analysis. And then a completely different experiment was suggested by Greenberger, Horn, and Zeilinger. I've got a reference here to a paper they wrote with Abner Shimoni\cite{greenberger-90ajp1131}, not because that was the original paper, but the original paper\cite{greenberger-89proc} is a brief report in conference proceedings. That one polishes it up. This is my version of Mermin's version\cite{mermin90ajp731,mermin90pt9} of Greenberger, Horn, and Zeilinger's\cite{greenberger-89proc} Gedanken experiment inspired by John Bell\cite{bell64physics195} based on Bohm\cite{Bohm51} and Einstein, Rosen, and Podolsky\cite{einstein-35pr777}. And I've left out ninety percent of the references.

The way I like to think of this analysis is by imagining a physicist, whom I call ``Dr.\ Diehard'', who was around at the time of the discovery of quantum mechanics in the late 20s, and didn't believe it. Although some time has passed since then, he's still around---quite old but intellectually vigorous, and he still doesn't believe in it. Our task is to convince him that quantum mechanics is right and classical ideas are wrong, or as I say even primitive pre-classical ideas.

There's no point in trying to wow him with the anomalous magnetic moment of the electron or the behavior of artificial atoms that we just heard about or anything like that, because he is so deeply opposed to quantum mechanics and so old and stubborn that as soon as you start putting a particular quantum mechanical equation on the board his brain turns off, rather like my brain in a seminar on string theory. So the only way to convince him is on very general grounds---not by doing particular calculations.

At first thought you say: ``It's easy---quantum mechanics is probabilistic, classical mechanics is deterministic. If I have that electron in an eigenstate of $\sigma_x$ and choose to measure $\sigma_z$, I can't tell whether I'm going to get $+1$ or $-1$. There's no way anyone can tell. That's very different from classical mechanics, and it seems to describe the real world''.
\begin{figure}[htb] 
\begin{mdframed}
  \vspace{5pt}\hfill 7\\ 
  \begin{raggedright}
    Dr.\ Diehard neither believes in nor understands quantum mechanics.  ``Deep down, it's all classical!''\\[\baselineskip]

    Probabilistic?  ``Just classical probability!''
    \begin{align}\nonumber
      A=A(&\alpha) 
    \end{align}
    where $\alpha =$ ``subquantum'' or ``hidden'' variables; 
    may be very many;
    may involve ``apparatus'' as well as ``system''.
    \begin{align}\nonumber
      \text{Prob}\{A\le a\}=\int \theta(a-A(\alpha)) d\mu(\alpha) 
    \end{align}
    where $\mu(\alpha) =$ probability distribution for the hidden variables---``a result of our ignorance not some quantum nonsense!'' \\[\baselineskip]

    Noncommuting Observables?  ``Just interfering measurements!''
  \end{raggedright}
  \vspace{6pt}
\end{mdframed}
\end{figure}

But Dr.\ Diehard is not convinced for a second by that. ``Probability has nothing to do with this fancy quantum mechanics. Jérôme Cardan was writing down the rules of probability when he analyzed games of chance in the late Renaissance. When I flip a coin or go to Las Vegas and have a spin on the roulette wheel, the results seem to be perfectly probabilistic. But I don't see Planck's constant playing any significant role there'', he says. ``The reason the roulette wheel gives me a probabilistic result is that there are all sorts of sensitive initial conditions which I can't measure well enough---initial conditions to which the final state of the ball is sensitive---there are all sorts of degrees of freedom of the system which I cannot control, and because of my ignorance, not because of any fundamental physics, I get a probabilistic
result.''

This is sometimes called the hidden variable position.  ``Really, you don't know everything about the state of the electron when you measure its momentum and its spin along the $x$-axis. There are zillions of unknown hidden variables which you can't control; maybe they are also in the system that is measuring the electron. (There's no separation in this viewpoint between the observer observing the system and the quantity being observed.) If you knew those quantities exactly, then you know exactly what the electron was going to do in any future experiment. But since you only know them probabilistically, you only have a probabilistic distribution.''

Here I've written it down in somewhat fancy-shmancy mathematical notation. [Coleman points at Slide 7.]  In fact, this is right---you can get probability from classical mechanics. John von Neumann way back was aware of this. He said: ``No, that's not the real difference between classical mechanics and quantum mechanics. The real difference is that in quantum mechanics you have non-commuting observables:  If you measure $\sigma_x$ repeatedly for an electron and take care to keep it isolated from the external world, you always get the same result.  But if you then measure $\sigma_z$ and get a probabilistic result, when you measure $\sigma_x$ again you will again get a probabilistic result the first time---the first measurement of $\sigma_z$ has interfered with the measurement of $\sigma_x$. That's because you have non-commuting observables, and those are characteristic of quantum mechanics.

Dr.\ Diehard says: ``Absolute nonsense! We're big, clumsy guys. When we think we're doing a nice clean measurement of $\sigma_x$ we might be messing up all of those hidden observables. When we measure your $\sigma_z$ we then get a different result because we've messed things up. My friends the anthropologists talk about this a lot when they discuss how an anthropologist can affect an isolated society he or she believes they're observing. And, for some reason I don't understand, they call it the uncertainty principle''.

And Dr.\ Diehard continues: ``My friends the social psychologists tell me that if you do an opinion survey, unless you construct it very carefully, the answers you will get to the questions will depend upon the order in which they are asked''. (This is true, by the way.) He doesn't see any difference between that and measurements of $\sigma_x$ and $\sigma_z$. That's Dr.\ Diehard's position.

As John Bell pointed out in the first written of those two articles I cited\cite{bell64physics195}---which is not the one with the famous inequality---this is in fact an irrefutable position, despite all the stuff to the contrary that has been said in the literature. On this level there is no way of refuting it. He gave a specific example of a classical theory that on this level reproduced all the results of quantum mechanics---the de Broglie pilot wave theory\cite{bohm52prb166}.

However, if Dr.\ Diehard admits one more thing, we can trap him. I will now explain what that one thing is.

\begin{figure}[htb] 
\begin{mdframed}
  \vspace{5pt}\hfill 8\\[-18pt]
  \begin{center}
    \pgfmathsetmacro{\sx}{2.5}
    \pgfmathsetmacro{\st}{2.2}
    \begin{tikzpicture}[>=latex]
      \begin{scope}[shift={(0,.5)}]
      \draw[->](-\sx,0)--(\sx,0) node[anchor=west]{$x$ (lt.yrs.)};
      \draw[->](0,-\st)--(0,\st) node[above]{$t$ (yrs.)};
      \draw [fill=black, thick] (0,0) circle[radius=.05] node[below left]{A};
      \draw [fill=black, thick] (1.7,.3) circle[radius=.05] node[above right]{B};
      \draw [fill=black, thick] (1.7,-.3) circle[radius=.05] node[below right]{B'};
    \end{scope}
    \end{tikzpicture}
  \end{center}
  \begin{raggedright}
    But spacelike-separated measurements can not interfere with each other (unless we have propagation of influence backward in time).\\[\baselineskip]

We have now a contradiction with the predictions of quantum mechanics for simple systems. 
  \end{raggedright}
  \vspace{6pt}
\end{mdframed}
\end{figure}

Here we have a drawing of space-time. It's really four dimensional, but due to budgetary constraints I've had to represent it as a two-dimensional object. The scale has been chosen so that time $t$ is measured in years and $x$ in light-years, therefore the paths of light rays are 45-degree lines.

Now let's consider two measurements on possibly two different systems done in two regions A and B---forget B' for the moment, its role will emerge later. Thus these black dots represent actually substantial regions in space time, during which an experiment has been conducted.

Now one thing Dr.\ Diehard will have to admit is that although the results of an experiment in A may interfere with an experiment in B, the results of an experiment in B can hardly interfere with the results of an experiment in A unless information can travel backwards in time, which we will assume he does not accept. That's because A is over and done with and its results recorded in the log book before B occurs.

On the other hand, if we imagine another Lorentz observer with another coordinate system, B will appear as B' here. B and B', as you can see by eyeball, are on the same space-like hyperbola---there is a Lorentz transformation that leaves A at the origin of coordinates unchanged and turns B into B'. B and B' are space-like separated from A. A light signal cannot get from A to B, and nothing traveling slower than the speed of light can get from A to B.

Now that second Lorentz observer would give the same argument I gave, except he would interchange the roles of A and B'. He would say the results of an experiment at A cannot interfere with the act of doing an experiment at B' because B' is earlier than A. But B' is B, just B seen by a different observer.

Therefore, if you believe in the principal of Lorentz invariance, and if you believe you cannot send information backwards in time, you have to conclude that experiments done at space-like separated locations sufficiently far apart from each other cannot interfere with each other. It can't matter what order you ask the questions if this question is being asked of an earthman and this one of an inhabitant of the Andromeda Nebula, and they're both being asked today.

Are there any questions about this? This is the groundwork from which the rest will proceed.
\begin{figure}[htb] 
\begin{mdframed}
  \vspace{5pt}\hfill 9\\[-11pt]
  \begin{center}
    \pgfmathsetmacro{\rad}{.5}
    \pgfmathsetmacro{\radi}{.4}
    \begin{tikzpicture}[>=latex]
      \draw[->](0,0)--(-1.59,-2.65);
      \draw[->](0,0)--(0,-3.2);
      \draw[->](0,0)--(1.59,-2.65);
      \draw [fill=white, thick](0,0) circle[radius=\rad] node{\large ?}; 
      \draw [fill=white, thick](-1.8,-3) circle[radius=\radi] node{1}; 
      \draw [fill=white, thick](0,-3.6) circle[radius=\radi] node{2}; 
      \draw [fill=white, thick](1.8,-3) circle[radius=\radi] node{3}; 
    \end{tikzpicture}
  \end{center}
  \begin{center}
    {Fig.~1: The Diehard Proposal (1'' = 1 light minute)}
  \end{center}
  \vspace{6pt}
\end{mdframed}
\end{figure}

On everything else we accept the Diehard position. Now here is the experimental proposal---this is a drawing from an imaginary proposal to the Department of Energy for the Diehard experiment. Three of Dr.\ Diehard's graduate students are assigned to experimental stations, as you see from the scale they are several light-minutes from each other. The graduate students, with lack of imagination, are called numbers 1, 2, and 3. They're almost as old as Dr.\ Diehard---it's difficult to get a thesis under him.

\begin{figure}[htb] 
\begin{mdframed}
  \vspace{5pt}\hfill 10\\[-20pt]
  \begin{center}
    \pgfmathsetmacro{\rad}{.75}
    \begin{tikzpicture}[>=latex ,scale=1]
      \draw [thick] (-2,-1.6) rectangle (2,1.6);
    \begin{scope}[shift={(-.8,0)}]
      \draw [very thick] (0,0)--(1.25,.72);
      \draw [fill=white, thick](0,0) circle[radius=.1]; 
      \draw [fill,thick](1.6,0) circle[radius=.06] node[right]{\ OFF}; 
      \draw [fill,thick](1.385,0.8) circle[radius=.06] node[right]{\ A}; 
      \draw [fill,thick](1.385,-0.8) circle[radius=.06] node[right]{\ B};
    \end{scope}
    \begin{scope}[shift={(1,1.6)}]
      \draw [thick] (-0.12,0) rectangle (.12,.6);
      \draw [fill=white, thick](0,.6) circle[radius=.35] node{$\bs{-}$}; 
    \end{scope}
    \begin{scope}[shift={(-1,2.2)}]
      \draw (0,0)--(0,\rad); 
      \draw[rotate=50] (0,0)--(0,\rad); 
      \draw[rotate=100] (0,0)--(0,\rad); 
      \draw[rotate=-50] (0,0)--(0,\rad); 
      \draw[rotate=-100] (0,0)--(0,\rad); 
      \draw [fill=white, white](0,0) circle[radius=.48];
      \draw [thick] (-0.12,-.6) rectangle (.12,0);
      \draw [fill=white, thick](0,0) circle[radius=.35] node{$\bs{+}$}; 
    \end{scope}
    \end{tikzpicture}
  \end{center}
  \vspace{-.4\baselineskip}
  \begin{center}
    {Fig.~2: The Acme ``Little Wonder'' Dual Cryptometer}
  \end{center}
  \vspace{6pt}
\end{mdframed}
\end{figure}
They are informed that once a minute something will be sent from a mysterious central station to each of the three Diehard teams---what something is, they don't know. However, they're armed with measuring devices whose structure they again do not know. They are called dual cryptometers because they can measure each of two things, but what those two things are nobody knows---at least the Diehards don't know. They can turn a switch to either measure $A$ or measure $B$. They make this decision once a minute shortly before the announced time of the signal, and sure enough, a light bulb lights up that says either $A$ is $+1$ or $A$ is $-1$ if they are measuring $A$, or the same thing for $B$. They have no idea what $A$ or $B$ is. It's possible the central station is sending them elementary particles. It's possible the central station is sending them blood samples, which they have the choice of analyzing for either high blood cholesterol or high blood glucose. It is possible the whole thing is a hoax, there is no central station, and a small digital computer inside the cryptometer is making the lights go on and off. They do not know.

\begin{figure}[htb] 
\begin{mdframed}
  \vspace{5pt}\hfill 11\\ 
  \begin{raggedright}
    The Diehard team obtains records like
  \end{raggedright}
\begin{align}\nonumber
  &&  A_1&=1 &B_2&=-1 &B_3&=-1&&\\\nonumber
  &&  A_1&=1 &A_2&=-1 &B_3&=-1&&\\\nonumber
  &&  B_1&=1 &B_2&=1  &A_3&=1 &&\\\nonumber
   &&&&&          \ldots   &&&&
\end{align}

\begin{raggedright}
  They find whenever they measure
  $A_1B_2B_3$ it is $+1$.\\
  Likewise for $B_1A_2B_3$ and $B_1B_2A_3$.\\[\baselineskip]
  They deduce that\\[-1.2\baselineskip]
\end{raggedright}
\begin{align}\nonumber
  \boxed{A_1A_2A_3=1} 
\end{align}
\vspace{-.5\baselineskip}
\end{mdframed}
\end{figure}
In this way, however, they obtain a sequence of measurements, which they record as this. [Coleman points to Slide 11.] The first line means observer 1 has decided to measure $A$ and obtained the result $+1$; observer 2 has decided to measure $B$ and obtained the result $-1$; and observer 3 has decided to measure $B$ and obtained the result $-1$. They have obtained in this way zillions of measurements on a long tape. They record them in this way because they really believe that whatever this thing is doing, $A_1=1$, that is to say, the value of quantity $A$ that would be measured at station 1 is $+1$ independent of what is going on on stations 2 and 3, because these three measurements are space-like separated. That's what they have to believe if they're Diehards. They have to believe there's really some predictable value of this thing which they would know if they knew all the hidden variables. In this particular case, they don't know what $B_1$ is but they know what $A_1$ is.

Now as they go through their measurements, they find in that roughly 3/8 of the measurements---they're making random decisions about which things they  measure---whenever they measure one $A$ and two $B$'s the result of the product of the measurements is $+1$. Now they're making their choices at random and since they believe that these things have well-defined meanings independent of their measurements, they have to believe, if they believe in normal empirical principles, that all the time the value of one $A$ and two $B$'s---the value that would be obtained if they had done the measurement---the product is $+1$. Sometimes all three of these numbers are $+1$. Sometimes one of them is $+1$ and two are $-1$. But the product is always $+1$. It's as if I gave you a zillion boxes and you turned up 3/8 of them and discovered each of them had a penny in it, you would assume within 1 over the square root of $N$---negligible error---that if you opened up all the other boxes, they would also have pennies in them. By the miracle of modern arithmetic---that is to say by multiplying these three numbers together and using the fact that each $B$ squared is 1---they deduce that if they look on their tape for those experiments in which they've chosen to measure the product of three $A$'s, they would obtain the answer $+1$.

\begin{figure}[htb] 
\begin{mdframed}
  \vspace{5pt}\hfill 12\\[-18pt]
  \begin{center}
    \textbf{Behind the Scenes}
  \end{center}
  \vspace{-5pt}
  \begin{align}\nonumber
    &\hspace{-20pt}
    \ket{\psi}=\frac{1}{\sqrt{2}}
    \bigl[\ket{\up\up\up} - \ket{\dw\dw\dw}\bigr]\\[5pt]\nonumber
    &\hspace{-20pt}
    A_1=\sigma_x^{(1)}\quad B_1=\sigma_y^{(1)} \quad\text{etc.}\\[5pt]\nonumber 
    &
    A_1B_2B_3\ket{\psi}
    =\sigma_x^{(1)}\sigma_y^{(2)}\sigma_y^{(3)}\ket{\psi}
    =\ket{\psi}\\[5pt]\nonumber 
    &
    \text{etc.\ for\ \ $B_1A_2B_3$\ \ and\ \ $B_1B_2A_3$.}\\[5pt]\nonumber 
    &\hspace{-20pt}\text{But \ldots}\\[0pt]\nonumber
    &
    A_1A_2A_3\ket{\psi}
    =\sigma_x^{(1)}\sigma_x^{(2)}\sigma_x^{(3)}\ket{\psi}
    =\bs{-}\ket{\psi}\\[5pt]\nonumber
    &\hspace{-20pt}\text{What spooky action-at-a-distance?}
  \end{align}
\vspace{-.5\baselineskip}
\end{mdframed}
\end{figure}
Now let's look behind the scenes and see what's actually going on.
Maybe a little suspense would help... [Coleman covers most of Slide 12].

It's not blood samples we're sending to them after all, it's three spin one-half particles arranged in the following peculiar initial state: one over the square root of two all spins up minus all spins down:
\begin{align}
  \frac{1}{\sqrt{2}}\bigl[\ket{\up\up\up} - \ket{\dw\dw\dw}\bigr]
\end{align}
$A$ is simply $\sigma_x$ for the particle that arrives at the appropriate station, and $B$ is $\sigma_y$.

Let's first check that $A_1B_2B_3$ acting on this state is $+1$.
By the third statement about quantum mechanics I put on the board in my preliminary section [see Slide 4], this quantity is definitely always going to be measured to be $+1$. Well, we have $\sigma_x(1)\sigma_y(2)\sigma_y(3)$ by my transcription table. $\sigma_x$ turns up into down. $\sigma_y$ turns up into down with a factor of $\i$ or maybe $-\i$, I can never remember,
but that's no problem here because you have two of them so the square is always $-1$.  Acting on the first component of this state this operator produces the second component including the minus sign while acting on the second component this operator produces the first. So this state is indeed an eigenstate of this operator with eigenvalue $+1$. And, of course, since everything is permutation invariant, the same is true for the other two operators.

But $A_1A_2A_3$ is $\sigma_x(1)\sigma_x(2)\sigma_x(3)$, and $\sigma_x$'s turns and up into a down without a minus sign. Therefore this state is also an eigenstate of $A_1A_2A_3$, but with eigenvalue $-1$.

The Diehards using only these proto-classical ideas---they aren't even so well developed to be called classical physics, they're sort of the underpinnings of classical reasoning---deduce that they will always get $A_1A_2A_3=+1$, sometimes a $+1$ and two $-1$'s, but always $+1$. In fact, if quantum mechanics is right, they will always get $-1$.

This is pedagogically superior to the original Bell argument for two reasons: Firstly, it doesn't involve correlation coefficients---it's not that classical mechanics says this will happen 47\% of the time and quantum mechanics says it happens 33\% of the time. Secondly, it is easy to remember---whenever I lecture on the Bell inequality I have to look it up again because I can never remember the derivation. This thing---the ingredients in it are so simple that if someone awakens you in the middle of the night four years from now, and puts a gun to your head, and says: ``show me the GHZM argument'', you should be able to do it.

We have shown that there are quantum mechanical experiments where the conclusions cannot be explained by classical mechanics---even the most general sense of classical mechanics---unless, of course, the classical mechanical person is willing to assume transmission of information faster than the speed of light, which, with the relativity principle, is tantamount to transmission of information backwards in time.

This is, of course, also John Bell's conclusion. This is, I must say, much misrepresented in the popular literature and even in some of the not so popular literature. That's not coming out right. I mean: some technical literature, where people talk about quantum mechanics necessarily implying connections between space-like separated regions of space and time. That's getting it absolutely backwards. There are no connections between space-like separated regions of space and time in this experiment. In fact, there's no interaction Hamiltonian, let alone one that transmits information faster than the speed of light, except maybe an interaction Hamiltonian between the individual cryptometers and the particles. But, otherwise, it's either quantum mechanics or superluminal transmission of information, not both.

Why on earth do people---I'm trying to see inside other people's heads, which is always a dangerous operation, but let me do it---why, why on earth do people get so confused, so wrong about such a simple point? Why do they write long books about quantum mechanics and non-locality full of funny arrows pointing in different directions? Okay, that's the technical philosophers. They really---well, I'll avoid the laws of libel---so, anyway, why do they do this? It's because, I think, secretly in their heart of hearts they believe it's really classical mechanics---that we're really putting something over on them---deep, deep down it's really classical mechanics.



\begin{figure}[htb] 
\begin{mdframed}
  \vspace{5pt}\hfill 12a\\ 
  \begin{raggedright}
    ``Every successful physical theory swallows its predecessor alive.''\\[\baselineskip]

    But it does so by interpreting the concepts of the old theory in terms of the new, NOT the other way around.\\[\baselineskip]

    Thus our aim is NOT ``the interpretation of quantum mechanics.''  It is the interpretation of classical mechanics.\\[\baselineskip]
  \end{raggedright}
\vspace{-.5\baselineskip}
\end{mdframed}
\end{figure}

People get things backwards and they shouldn't---it has been said, and wisely said, that every successful physical theory swallows its predecessors alive. By that we mean that in the appropriate domain---for example the way statistical mechanics swallowed thermodynamics---in the appropriate domain of experience, the fundamental concepts of thermodynamics---entropy for example, or heat---were explained in terms of molecular motions, and then we showed that if you defined heat in terms of molecular motion it acted under appropriate conditions pretty much the way it acted in thermodynamics. It's not the other way around. The thing you want to do is not to interpret the new theory in terms of the old, but the old theory in terms of the new.

The other day I was looking at a British videotape of Feynman explaining elementary concepts in science to an interrogator, whom I think was the producer 
Christopher Sykes.  He asked Feynman to explain the force between magnets.
Feynman hemmed and hawed for a while, and then he got on the right track, and he said something that's dead on the nail. He said: ``You've got it all backwards, because you're not asking me to explain the force between 
your pants and the seat of your chair.  You want me, when you say the force between magnets, to explain the force between magnets in terms of the kinds of forces you think of as being fundamental---those between bodies in contact''. Obviously, I'm not phrasing it as wonderfully as Feynman. But, well, as Picasso said in other circumstances, it doesn't have to be a masterpiece for you to get the idea. We physicists all know it's the other way around: the fundamental force between atoms is the electromagnetic force which does fall off as one over $R$ squared. Christopher Sykes was confused because he was asking something impossible.  He should have asked to explain the pants-chair force in terms of the force between magnets.
%
Instead he asked to
derive the fundamental quantity in terms of the derived one.

Likewise, a similar error is being made here. The problem is not the interpretation of quantum mechanics. That's getting things just backwards. The problem is the interpretation of classical mechanics.

Now, I'm going to address this, and in particular the famous, or infamous, projection postulate.

\begin{figure}[htb] 
\begin{mdframed}
  \vspace{5pt}\hfill 13\\[-18pt]
  \begin{center}
    \textbf{(3) Credits for the next part}
  \end{center}
  \begin{raggedright}
  \begin{enumerate}[label={[\roman*]}]\setcounter{enumi}{5}
    \item J. von Neumann, \emph{Mathematische Grundlagen der Quantenmechanik} (1932).
    \item H. Everett (1957), 
      Rev. Mod. Phys. 29, 454 (1957).
    \item J. Hartle, 
      Am. J. Phys. 36 (1968) 704.
    \item E. Farhi, J. Goldstone, and S. Gutmann, 
      Ann. Phys. (NY) 192, (1989) 368.

  \end{enumerate}
  \end{raggedright}
  \vspace{6pt}
\end{mdframed}
\end{figure}

The fundamental analysis is von Neumann's. I don't read two words of German, but I wanted to put down this early publication\cite{Neumann32}. I read it in English translation\cite{Neumann55}. The position I am going to advocate is associated with Hugh Everett in a classic paper\cite{everett57rmp454}. Some of the things I'll say about probability later come from a paper by Jim Hartle\cite{hartle68ajp704}, and one by Cambridge's own Eddie Farhi, Jeffrey Goldstone, and Sam Gutmann\cite{farhi-89ap368}.

I'd like to begin by recapitulating von Neumann's analysis of the measurement chain.

\begin{figure}[htb] 
\begin{mdframed}
  \vspace{5pt}\hfill 14\\[-18pt]
  \begin{center}
    \textbf{The Measurement Chain} (after von Neuman)
  \end{center}
  \begin{raggedright}
  \begin{enumerate}[label={(\arabic*)}]\setcounter{enumi}{0}
    \item Electron prepared in $\sigma_\x$ eigenstate:
      \begin{align}\nonumber
        &\hspace{-20pt}
          \ket{\psi}=\frac{1}{\sqrt{2}}
          \bigl[\ket{\up} + \ket{\dw}\bigr]
      \end{align}
      I measure $\sigma_\z$:
      \begin{align}\nonumber
        &
          \ket{\psi}=
          \begin{cases}
            \ket{\up} \\[3pt] \ket{\dw}
          \end{cases}
        \text{equal probabilities}
      \end{align}
      Non-deterministic ``reduction of the wave function''
    \item Electron as before, measuring device in ground state:
      \begin{align}\nonumber
        &\hspace{-20pt}
          \ket{\psi}=\frac{1}{\sqrt{2}}
          \bigl[\ket{\up,\M_0} + \ket{\dw,\M_0}\bigr]
      \end{align}
      Electron interacts with the device:
      \begin{align}\nonumber
        &\hspace{-20pt}
          \ket{\psi}\rightarrow\frac{1}{\sqrt{2}}
          \bigl[\ket{\up,\M_+} + \ket{\dw,\M_-}\bigr]
      \end{align}
      (normal deterministic time evolution)\\[5pt]
      I observe device:
      \begin{align}\nonumber
        &
          \ket{\psi}=
          \begin{cases}
            \ket{\up,\M_+} \\[3pt] \ket{\dw,\M_-}
          \end{cases}
        \text{equal probabilities}
      \end{align}
  \end{enumerate}
  \end{raggedright}
\vspace{-.5\baselineskip}
\end{mdframed}
\end{figure}

I prepare an electron in a $\sigma_x$ eigenstate and I measure $\sigma_z$---the famous non-deterministic ``reduction of the wave packet'' takes place, and with equal probabilities, I cannot tell which, the spin either goes up or down.

But this is rather unrealistic even for a highly idealized measurement. An electron is a little tiny thing, and I have bad eyes. I probably won't be able to see directly what its spin is. There has to be an intervening measuring device. So we complicate the system.

The initial state is the same as before, as far as the electron goes, but the measuring device is in some neutral state [M$_0$ on Slide 14]%
. The electron interacts with the measuring device. Von Neumann showed us how to set things up with the interaction Hamiltonian so if the electron is spinning up the measuring device goes---maybe it's one of those dual cryptometers---the light bulb saying $+1$ flashes, if the electron is spinning down the light bulb saying $-1$ flashes. This is normal deterministic time evolution according to Schrödinger's equation.

Now I come by. I can't see the electron, but I observe the device. By the usual projection postulate, I either see it in state $+1$ or state $-1$. I make the observation, if I see the state $+1$, and the rest of the wave function is annihilated. I get with either probability these two things [Coleman points at the state vectors $\ket{\up,M_+}$ and $\ket{\dw,M_-}$ on the bottom of Slide 14].  The result is the same as before because the electron is entangled with the device. I measure the device. The electron comes along for the ride.

Now let's complicate things a bit more. Let's suppose I cannot do the measurement because I'm giving this lecture. However, I have a colleague, a very clever experimentalist---for purposes of definiteness, let's say its Paul Horowitz---who has constructed an ingenious robot. I'll call him Gort. It's a good name for a robot. I say ``Gort, I want you during the lecture to go and see what the measuring device says about the electron''. And so Gort goes and does this. Although he's an extremely ingenious and complicated robot, he's still just a big quantum mechanical system, like anything else. So it's the same story. Things starts out with electron in a superposition of up and down, 
measuring device neutral, a certain register and a RAM chip inside Gort's belly also has nothing written on it. Then everything interacts and the state of this world is: electron ``up'', measuring devices says ``up'', Gort's RAM chip's register says ``up'', plus the same thing with ``up'' replaced by ``down'', all divided by the square root of two. And Gort comes rolling in the door there with his rollers, and I say: ``Hey, Gort, which way is the electron spinning?'' And he tells me. And wham-o, it either goes into one or the other of these states fifty percent probability.

But Gort is very polite. He observes that I am lecturing. So rather than coming to me directly, he rolls up to my colleague Professor Nelson sitting there in the corner and hands him a clip of printout that says either up or down, and says: ``Pass this on to Sidney when the lecture is over''. And he rolls away.

Well, of course, vitalism was an intellectually live position early in the 19th century. Dr.\ Lydgate in Middlemarch, which will be appearing on TV tomorrow, held that living creatures are not simply complicated mechanical systems. But it hasn't had many advocates this century. I think most of us would admit that David [Nelson] is just another quantum mechanical system, although perhaps more complicated than the electron and Gort, and certainly more likeable. Anyway, there he is.

So it's the same story as before: the state of the world after all this has happened is: electron ``up'', measuring device says ``up'', Gort's RAM chip says ``up'', David's slip of paper says ``up'', plus the same thing with down, divided by the square root of two. After the lecture I go up to them and say: ``What's up, David?'' Wham-o! He tells me. And the whole wave function collapses.

Now this is getting a little silly, especially if you consider the possibility that---after all, I'm getting on in years, I'm not in perfect health, here I am running around a lot---maybe I have a heart attack before the lecture is over and die. What happens then? Who reduces the wave packet?

Yakir Aharonov, who has of course since acquired great fame for himself, was a young postdoc at Brandeis when I was a young postdoc at Harvard. I had been reading von Neumann and thinking about this, and come to a conclusion which I did not like, which was solipsism: I was the only creature in the world which could reduce wave packets. Otherwise it didn't make any sense. I was not totally happy with this position, even though I was as egotistical as any young man---indeed probably more egotistical than most---I was still unhappy with the position. I was discussing this with Aharonov. Even in his youth he would smoke these enormous cigars, which he would use to punctuate the conversation; he would take huge drafts on them; he was and is sort of the quantum George Burns.

Anyway, I explained this position to him, and he said: ``I see. [Coleman imitates Aharonov inhaling and blowing out smoke of his cigar] Tell me: before you were born, could your father reduce wave packets?''

\begin{figure}[htb] 
\begin{mdframed}
  \vspace{5pt}\hfill 15\\[-18pt]
  \begin{raggedright}
  \begin{enumerate}[label={(\arabic*)}]\setcounter{enumi}{2}
    \item Add robot
    \item Add colleague\\[5pt]
  \end{enumerate}

    The problem of death. 
    Aharonov's question.\\[\baselineskip]
    I will argue the there is
    \begin{itemize}[label={}]
    \item {\bf NO}\ \ special measurement process  
    \item {\bf NO}\ \ reduction of the wave function 
    \item {\bf NO}\ \ indeterminancy
    \item {\bf NOTHING}\ \ probabilistic
    \end{itemize}
    in quantum mechanics.
    \begin{itemize}[label={}]
    \item {\bf ONLY}\ \ deterministic evolution\\ 
      \hspace{33pt} according to Schrödinger's Equation\\[10pt]
    \end{itemize}

    ``Ridiculous''---E.~Schrödinger (1935) \\[0.5\baselineskip]

    ``Absurd''---E.P.~Wigner (1961) \\[0.5\baselineskip]

    ``Why do I, the observer, perceive only one of the\\ outcomes?''---W.H.~Zurek (1991)
  \end{raggedright}

\vspace{0.5\baselineskip}
\end{mdframed}
\end{figure}

Now I will argue that in fact there is no special measurement process, there is no reduction of the wave function in quantum mechanics, there is no indeterminacy, and nothing probabilistic---only deterministic evolution according to Schrödinger's equation.

This is not a novel position.  In the famous paper on the cat, Schrödinger\cite{schroedinger35naturwissenschaften807} raised this position, the position that the cat is in fact in the coherent superposition of being dead and being alive, and instantly said it's ridiculous: ``We reject the ridiculous possibility ...''

Some years later in the paper on Wigner's friend, where Wigner\cite{Wigner61} attempted to resolve the ancient mind-body problem through the quantum theory of measurement, he also raised this position, and said it was ``absurd''.

There is a recent paper by Zurek \cite{zurek91pt9} in \emph{Physics Today}---Zurek has made major contributions to the theory of decoherence
---where instead of just saying it's ridiculous or absurd, he actually raised a question one can talk about. He said: ``If this is so, why do I the observer perceive only one of the outcomes?'' This is now the question I will attempt to address: Zurek's question. If there is no reduction of the wave packet, why do I feel at the end of the day that I have observed a definite outcome, that the electron is spinning up or the electron is spinning down?

\begin{figure}[htb] 
\begin{mdframed}
  \vspace{5pt}\hfill 15a\\[3pt] 
  \begin{raggedright}
      N.\ Mott (1929) asked: ``If an ionized particle is emitted in
      an s-wave state in the center of a cloud chamber, why is the
      ionization track a straight line rather than some spherical
      symmetric distribution?''  [Of course, we must assume that
        particle momentum is unchanged (to within some small angle)
        when it scatters off an atom.]\\[5pt]
      Let $\ket{\text{C}}$ be the state of the cloud chamber.\\[5pt]
      Define a ``linearity operator'' $L$, such that   
  \begin{align}\nonumber
    &
    L\ket{\text{C}}=\ket{\text{C}}
    &&\text{if track is straight (t.w.s.s.a.)},\\[2pt]\nonumber
    &
    L\ket{\text{C}} = 0 &&\text{on states orthogonal to these.}
  \end{align}
  \vspace{-15pt}
  \begin{align}\nonumber
    &\hspace{-20pt}\ket{\psi_\i}=\ket{\phi_{\bs{k}},\text{C}_0}\,\to\,
    \ket{\psi_{\text{f},\bs{k}}}
  \end{align}
  where $\phi_{\bs{k}} =$ state where the particle is concentrated near
  the center in position and near $\bs{k}$ in momentum.
  \begin{align}\nonumber
    &\hspace{-20pt} L\ket{\psi_{\text{f},\bs{k}}}=\ket{\psi_{\text{f},\bs{k}}}
  \end{align}
  Now consider:
  \begin{align}\nonumber
    &\hspace{-20pt}\ket{\psi_\i}
    =\int d\Omega_{\bs{k}}\ket{\phi_{\bs{k}},\text{C}_0}\,\to\,
    \ket{\psi_{\text{f}}}=\int d\Omega_{\bs{k}}\ket{\psi_{\text{f},\bs{k}}}
  \end{align}
  \vspace{-15pt}
  \begin{align}\nonumber
    &\hspace{-20pt} L\ket{\psi_{\text{f}}}=\ket{\psi_{\text{f}}}
  \end{align}
  \end{raggedright}
\vspace{-.5\baselineskip}
\end{mdframed}
\end{figure}

In order to ease into this, I'd like to begin with an analysis of Neville Mott\cite{mott29prsla79}. Neville Mott worried way back in 1929 about cloud chambers. He said: ``Look, an atom releases an ionizing particle at the center of a cloud chamber in an s-wave. And it makes a straight line track. Why should it make a straight line track? If I think about an s-wave, it is spherically symmetric. Why do they not get some spherically symmetric random distribution of sprinkles? Why should the track be a straight line?''

Now we're going to answer that question, and in a faster and slicker way then Neville Mott did. Of course, we have the advantage of 65 years of hindsight.

We must assume that the scattering between the particle and an atom when it ionizes it is unchanged or changed only within some small angle to begin with; otherwise, of course, even classically the particle would bounce around like a pinball on a pinball table.

Let $\ket{\text{C}}$ be the state of the cloud chamber. We define a linearity operator $L$---a projection operator so that
$L$ on $\ket{\text{C}}$ equals $\ket{\text{C}}$
if there is a track and it forms a straight line to within some small angle, and $L$ on $\ket{\text{C}}$ equals zero
if the track is not a straight line, or there is no track for that matter.
Now let's imagine we start out the problem in some initial state where the particle is concentrated near the center of the chamber and near some momentum $\bs{k}$, and the cloud chamber in a neutral condition, all unionized ready to make tracks. This evolves into some final state.

Now we all believe that if you started out with the particle in a narrow beam it would of course make a straight line track along that beam. The final state would be an eigenstate of this linearity operator and would have eigenvalue $+1$.

Now here comes the tricky part: not tricky to follow but tricky-clever. I consider an initial state that's an integral over the angles of $\bs{k}$ of this state. This is a state where the particle is initially in an s-wave, and the cloud chamber is still in a neutral state---that's independent of $\bs{k}$. That state evolves by the linearity---the causal linearity of Schrödinger's equation---into the corresponding superposition of these final states here [Coleman points to the state
$\ket{\psi_{\text{f}}}$ on the bottom of Slide 15a]. But if I have a linear superposition of eigenstates of the particle with respect to the operator $L$, each of which is an eigenstate with eigenvalue $+1$, then the combination is also an eigenstate with eigenvalue $+1$. So this also has straight line tracks in it.

That's the short version of Mott's argument. Mott said the problem is that people think of the Schrödinger equation as a wave in a three-dimensional space rather than a wave in a multi-dimensional space. I would phrase that, making a gloss on this---he's dead, so I can't check whether it's an accurate phrasing---I would make a gloss on this and say: the problem is that people think of the particle as a quantum mechanical system but of the cloud chamber as a classical mechanical system. If you're willing to realize that both the particle and the cloud chamber are two interacting parts of one quantum mechanical system, then there's no problem. It's an s-wave not because the tracks are not straight lines but because there is a rotationally invariant correlation between the momentum of the particle and where the straight line points. But it's always an eigenstate of this linearity operator.

Any questions about this?

Nobody doubts it---the tracks in cloud chambers, or bubble chambers if you're young enough, are straight lines, even if the initial state is an s-wave.
\begin{figure}[htb] 
\begin{mdframed}
  \vspace{5pt}\hfill 16\\[-5pt] 
  \begin{raggedright}
    To answer Zurek's question, we must assume a (quantum) mechanical theory of consciousness.
    \begin{align}\nonumber
        &\hspace{-20pt}
          \ket{\S}\in \mathcal{H}_{\S}
          &&\hspace{-20pt}\text{Hilbert space of 
             states}\\[-3pt]\nonumber
        & &&\hspace{-20pt}\text{of observer (Sidney)}
    \end{align}
  Introduce $D$, ``the definiteness operator'':
  \begin{align}\nonumber
    &\hspace{0pt}
      D\ket{\S}=\ket{\S}
    &&\hspace{-10pt}\text{if observer feels he has 
       perceived}\\[-3pt]\nonumber
    & &&\hspace{-10pt}\text{only one of the
         outcomes}\\[0pt]\nonumber
    &\hspace{0pt}
      D\ket{\S}=0
    &&\hspace{-10pt}\text{on states orthogonal to these.}     
  \end{align}
  \begin{align}\nonumber
    (1)\hspace{10pt} \ket{\psi_\i}&=\ket{\up,\M_0,\S_0}\ \to \ 
             \ket{\psi_\f}=\ket{\up,\M_+,\S_+}\\[3pt]\nonumber
    &\hspace{33pt} D\ket{\psi_\f}=\ket{\psi_\f}\\[6pt]\nonumber
    (2)\hspace{10pt} \ket{\psi_\i}&=\ket{\dw,\M_0,\S_0}\ \to \ 
             \ket{\psi_\f}=\ket{\dw,\M_-,\S_-}\\[3pt]\nonumber
    &\hspace{33pt} D\ket{\psi_\f}=\ket{\psi_\f}\\[6pt]\nonumber
    (3)\hspace{10pt} \ket{\psi_\i}&=\frac{1}{\sqrt{2}}
    \bigl[\ket{\up,\M_0,\S_0}
                     +\ket{\dw,\M_0,\S_0}\bigr]
    \\[1pt]\nonumber
    \ \to\ \ket{\psi_\f}&=\frac{1}{\sqrt{2}}
    \bigl[\ket{\up,\M_+,\S_+}+\ket{\dw,\M_-,\S_-}\bigr]
    \\[3pt]\nonumber
    &\hspace{33pt} D\ket{\psi_\f}=\ket{\psi_\f}
  \end{align}
 \end{raggedright}
\vspace{-.75\baselineskip}
\end{mdframed}
\end{figure}

Now I will give an argument due to David Albert\cite{Albert94} with respect to Zurek's question. Zurek asked: ``Why do I always have the perception that I have observed a definite outcome?'' To answer this question, no cheating: we can't assume Zurek is some vitalistic spirit loaded with \emph{élan vital} unobeying the laws of quantum mechanics. We have to say the observer---well I don't want to make it Zurek, that would be using him without his permission, I'll make it me, Sidney---has some Hilbert space of states, and some condition in Sidney's consciousness corresponds to the perception that he has observed a definite outcome, so there is some projection operator on it, the definiteness operator.
If you want, we could give it an operational definition: the state
where the definiteness operator is $+1$ is one where a hypothetical polite interrogator asks Sidney: ``Have you observed a definite outcome?'', and he says: ``Yes''. In the orthogonal states he would say: ``No, gee, I was looking someplace else when that sign flashed'' or ``I forgot'' or ``Don't bother me, man, I'm stoned out of my mind'' or, you know, any of those things.

Now let's begin. Our same old system as before: electron, measuring apparatus, and Sidney. If the electron is spinning up, the measuring apparatus measures spin in the up direction, and we get a definite state---no problem of superposition---and Sidney thinks: ``I've observed a definite outcome''. Same if everything is down. What if we start out with a superposition? Same story as Neville Mott's cloud chamber. The same reason the cloud chamber always shows the track to be a straight line is the reason Sidney always has the feeling he has observed a definite outcome.


[Coleman answers an inaudible question:]
That's not what Zurek said. Zurek didn't say: ``It's a matter of common experience that in this experiment we always observe the electron spinning up'', and Neville Mott didn't say: ``It's a matter of common experience that in the cloud chamber the straight line is always pointing along the z-axis''. The matter of common experience is that Sidney always has the perception that he has observed a definite outcome if you set up the initial conditions correctly. The matter of common experience is that the cloud chamber is always in a straight line. If you don't like this argument [the argument why Sidney perceives definite outcomes], you can't like that one [the argument why the cloud chamber detects straight lines]. If you like that one, you have to like this one.

The problem there---the confusion Nevil Mott removed---was refusing to think of the cloud chamber as a quantum mechanical system. The problem here is refusing to think of Sidney as a quantum mechanical system.

Because of the pressure of time I will remove these transparencies now and go on to discuss the question of probability.

\begin{figure}[htb] 
\begin{mdframed}
  \vspace{5pt}\hfill 17\\[-18pt]
  \begin{center}
    \textbf{What about probability?} 
  \end{center}
  \begin{raggedright}
    Classical probability theory reviewed:

    Suppose we have an infinite sequence of coin flips, or, equivalently, a sequence $\sigma_r (r=1,2,\ldots )$ of plus and minus ones.  We have a sequence of independent random flips of a fair coin if
    \begin{align}\nonumber
      &\hspace{-20pt}
        \bar\sigma =\lim_{N\to\infty}\bar\sigma^N=
        \lim_{N\to\infty}\frac{1}{N}\sum_{r=1}^N\sigma_r=0
      \\[-18pt]\nonumber
    \end{align}
    and\vspace{-5pt}
    \begin{align}\nonumber
      &\hspace{-18pt}
        \bar\sigma^a =\lim_{N\to\infty}\bar\sigma^{N,a}=
        \lim_{N\to\infty}\frac{1}{N}\sum_{r=1}^N\sigma_r\sigma_{r+a}=0
    \end{align}
    for all $a$. Also
    \begin{align}\nonumber
      &\hspace{-20pt}
        \lim_{N\to\infty}\frac{1}{N}
        \sum_{r=1}^N\sigma_r\sigma_{r+a}\sigma_{r+b}=0
    \end{align}
    for all $a,b$. Etc.

    (some mathematical niceties ignored.)
  \end{raggedright}
\vspace{.5\baselineskip}
\end{mdframed}
\end{figure}

Probability is a difficult question to discuss because it requires us to look at something counterfactual. 
If I ask whether a given sequence is or is not random, I can't do that even in classical probability theory for a finite sequence. For example, if I consider a binary sequence where the entries are either $+1$ or $-1$, and ask whether the sequence $+1$ is a random sequence, obviously there is no way of answering that question. But if I have an infinite sequence I can ask whether it's random. So let me talk about that.

Let me suppose I have an infinite sequence of $+1$ and $-1$'s, which might represent heads and tails. I want to see if these sequences can be interpreted as a fair coin flip. Firstly, I want the average value of this quantity $\sigma_r$, which is of course the limit of the average of the first $N$ terms  as $N$ goes to infinity, to converge to zero. Also, if I were an experimenter, I would probably look at correlations. I would take
the $r$-th value $\sigma_r$ times the $(r\hspace{-2pt}+\hspace{-2pt}a)-$th value $\sigma_{r+a}$ for some value of $a$, and look at the limit of this correlation, and ask that this quantity be also 0 for any value of a. That way I could reject sequences like $+1$,$+1$,$-1$,$-1$,$+1$,$+1$,$-1$,$-1$ \ldots, which no one would call random.  I could also look for triple and higher correlations. And if all those things were zero then I say there is a pretty good chance of a random sequence.

I would actually have to provide even further tests if I wanted the real definition of randomness, the Martin-L\"of definition of randomness\cite{martin-lof66ic602}, but this will be good enough for a lecture where I only have five minutes left.

\begin{figure}[htb] 
\begin{mdframed}
  \vspace{5pt}\hfill 18\\[-18pt]
  \begin{raggedright}
  \begin{align}\nonumber
    &\hspace{-20pt}
      \ket{\to }\equiv
      \frac{1}{\sqrt{2}}\bigl(\ket{\up}+\ket{\dw}\bigr)      
      \\[-18pt]\nonumber
  \end{align}
  Consider
  \begin{align}\nonumber
    &\hspace{-20pt}
     \ket{\psi}
      =\ket{\to}\otimes\ket{\to}\otimes\ket{\to}\cdots      
  \end{align}
  This is an infinite sequence of electrons, each with $\sigma_\x=1$. Let these interact with a $\sigma_z$-measuring device and an observer, as before.  Does the observer perceive a sequence of independent random flips?

  Equivalently, is $\ket{\psi}$ an eigenstate of
  \begin{align}\nonumber
    &\hspace{-20pt}
      \bar\sigma_\z =\lim_{N\to\infty}\bar\sigma_\z^N=
      \lim_{N\to\infty}\frac{1}{N}\sum_{r=1}^N\sigma_\z^{(r)}=0
  \end{align}
  with eigenvalue zero?  (And likewise for $\sigma_\z^a$ etc.)
  \end{raggedright}
\vspace{.5\baselineskip}
\end{mdframed}
\end{figure}

Now we want to ask the parallel question in quantum mechanics. We start out with an electron in the state I'll call sidewise---just our good old $\sigma_x$-eigenstate, the same state I've used before. I consider an infinite sequence of electrons heading towards my $\sigma_z$-measuring apparatus, and I do the usual routine with the measuring system in Sidney's head and turn it into a sequence of memories in Sidney's head or maybe Sidney has a notebook and he writes down $+1$,$-1$,$+1$,$+1$,$-1$. I obtain a sequence of records correlated with the z-component of spin. I ask: ``Does this observer observe this as a random sequence? That is to say, is this state here an eigenstate of the corresponding quantum observables with eigenvalue zero?''

Well, we know it's all correlated with $\sigma_z$. In order to keep the transparency from overflowing its boundaries, I just looked at $\sigma_z$, rather than the operator, for the records. I define the average value
of $\sigma_z$ exactly the same way as it is done up here.  [Coleman points to the lower equation of Slide 18.]  Then I ask:  
Is this an eigenstate of this operator with eigenvalue zero? If it is, we can say---despite the fact that there is nothing probabilistic in here---that the average value of $\sigma_z$ is guaranteed to be observed to be zero.

\begin{figure}[htb] 
\begin{mdframed}
  \vspace{5pt}\hfill 19\\[-18pt]
  \begin{raggedright}
  \begin{align}\nonumber
    &\bignorm{\,\bar\sigma_\z^N\ket{\psi}\, }^2
      =\frac{1}{N^2}\bra{\psi}
      \sum_{r=1}^N \sum_{s=1}^N\sigma_\z^{(r)}\sigma_\z^{(s)}\ket{\psi}
    \\[3pt]\nonumber
    &
      \bra{\psi}\sigma_\z^{(r)}\sigma_\z^{(s)}\ket{\psi}=\delta^{rs}
    \\[6pt]\nonumber
    &\Rightarrow\quad
    \lim_{N\to\infty}\bignorm{\,\bar\sigma_\z^N\ket{\psi}\, }^2
    =\lim_{N\to\infty}\frac{1}{N^2} N =0
  \end{align}
  Likewise for $\sigma_\z^a$ etc. \\[0.5\baselineskip]

  A definite deterministic state, definitely a random sequence.
  (An impossibility in classical physics---but this is not classical physics.)\\[\baselineskip]

  Stoppard's Wittgenstein.
  \end{raggedright}
\vspace{.5\baselineskip}
\end{mdframed}
\end{figure}

Well, the calculation is sort of trivial. Let's compute the norm of the state obtained by applying this operator to this state.  [Coleman points to the lower equation of Slide 19.]  It's two sums, and here I've written them out. Each of them is an individual thing, there's a 1 over $N$-squared, there's a sum on $r$, and a sum on $s$. Now in this particular state of course if $r$ is not equal to $s$ this ``expectation value'' is equal to zero, because you get just the product of the independent expectation values which are individually zero. On the other hand, if $r$ is equal to $s$, then this is $\sigma_z$ squared, which we all know is $+1$. Therefore, the limit of this thing up here is the limit of 1 over $N$ squared---the double sum collapses to a single sum, only the terms with $r$ equals $s$ contribute, and each entry with $r$ equals $s$ contributes 1, so you get $N$. Thus the result is $N$ over $N$ squared, which is of course 0.

And the same thing happens for all those correlators, because each one is a sum of terms with a 1 over $N$ squared in front and only the entries that match perfectly will give you a nonzero contribution. So this definitely quantum mechanical state completely determined by the initial conditions nevertheless matches this experimenter's definition of randomness---something that would be impossible in classical mechanics, but it's quantum mechanics, stupid.

Now one final remark:
In Tom Stoppard's play \emph{Jumpers}, there's an anecdote about the philosopher Ludwig Wittgenstein. I have no idea whether it's a real story or a Cambridge folk story\cite{stoppard}. Anyway, it goes like this.  A friend is walking down the street in Cambridge and sees Wittgenstein standing on a street corner lost in thought, and said: ``What's bothering you, Ludwig?'' Wittgenstein says: ``I was just wondering why people said it was natural to believe the sun went around the earth rather than the other way around''. The friend says: ``Well, that's because it looks like the Sun goes around the earth''. Wittgenstein thinks for a moment and says: ``Tell me: What would it have looked like if it had looked like it was the other way around?''



Now people say the reduction of the wave packet occurs because it looks like the reduction of the wave packet occurs, and that is indeed true. What I'm asking you in the second main part of this lecture is to consider seriously what it would look like if it were the other way around---if all that ever happened was causal evolution according to quantum mechanics. What I have tried to convince you is that what it looks like is ordinary everyday life. Welcome home. Thank you for your patience.


\vfill\eject\vfill 

\begin{acknowledgments}
MG wishes to thank Diana Coleman granting permission to publish this writeup, and Tobias Helbig for his critical reading of the manuscript.  The editorial effort 
was supported by the Deutsche Forschungsgemeinschaft (DFG, German Research Foundation)---Project-ID 258499086---SFB 1170 and through the W\"urzburg-Dresden Cluster of Excellence on Complexity and Topology in Quantum Matter---\textit{ct.qmat} Project-ID 390858490---EXC 2147.
\end{acknowledgments}



\end{document}